# Evidence for the novel type of orbital Fulde-Ferrell-Larkin-Ovchinnikov state in the bulk limit of 2H-NbSe$_2$

Chang-woo Cho[1,2], Timothée T. Lortz[1], Kwan To Lo[1], Cheuk Yin Ng[1], Shek Hei Chui[1], Abdel Rahman Allan[3], Mahmoud Abdel-Hafiez[3,4], Jaemun Park[5], Beopgil Cho[5], Keeseong Park[5], Noah F. Q. Yuan[6], and Rolf Lortz[1]

[1]Department of Physics, The Hong Kong University of Science and Technology, Clear Water Bay, Kowloon, Hong Kong, China

[2]Department of Physics, Chungnam National University, Daejeon, 34134, Republic of Korea

[3]Department of Applied Physics and Astronomy, University of Sharjah, P. O. Box 27272 Sharjah, United Arab Emirates

[4]Department of Physics and Astronomy, Uppsala University, Uppsala SE-75120, Sweden

[5]Department of Physics and Chemistry, Daegu Gyeongbuk Institute of Science and Technology (DGIST), Daegu, 42988, Republic of Korea

[6] Tsung-Dao Lee Institute and School of Physics and Astronomy, Shanghai Jiao Tong University, Shanghai 200240, China

## Abstract

The Fulde-Ferrell-Larkin-Ovchinnikov (FFLO) state, an unusual superconducting state, defies high magnetic fields beyond the Pauli paramagnetic limit. It exhibits a spatial modulation of the superconducting order parameter in real space and is exceptionally rare. Recently, an even more exotic variant—the orbital FFLO state—was predicted and identified in the transition metal dichalcogenide superconductor 2H-NbSe$_2$. This state emerges in thin samples with thicknesses below ~40 nm, at the boundary between two and three dimensions. The complex interplay between Ising spin orbit coupling and the Pauli paramagnetic effect can lead to a stabilization of the FFLO state in a relatively large range of the magnetic phase diagram, even well below the Pauli limit. In this study, we present experimental evidence of the formation of this orbital FFLO state in bulk 2H-NbSe$_2$ samples. This evidence was obtained using high-resolution DC magnetization and magnetic torque experiments in magnetic fields applied strictly parallel to the NbSe$_2$ basal plane. Both quantities display a crossover to a discontinuous first-order superconducting transition at the normal state boundary in magnetic fields of 3 T and above, which vanishes for small misalignments of the field. This is usually seen as a sign that Pauli paramagnetic pair breaking effects affect the superconducting state. Both the magnetic torque and the DC magnetization reveal a small step-like reversible anomaly, indicating a thermodynamic phase transition within the superconducting state. This anomaly bears many similarities to the FFLO transitions in other FFLO superconductors, suggesting the potential existence of an orbital FFLO state in bulk 2H-NbSe$_2$ samples. Additionally, we observe a pronounced in-plane 6-fold symmetry of the upper critical field in the field range above this phase transition, which has previously been interpreted as a hallmark of the orbital FFLO state in thin 2H-NbSe$_2$.

## Introduction

Superconducting transition metal dichalcogenides (TMDs) with their strongly anisotropic layered structures exhibit remarkable resilience against strong magnetic fields, especially when these fields align parallel to their layered structure. This resilience manifests differently in two-dimensional (2D) and three-dimensional (3D) forms of TMDs. In the 2D form, the unique combination of an open Fermi surface and a specific type of Ising spin orbital coupling (SOC) neutralizes the conventional upper critical field mechanism [1-3]. On the other hand, in the 3D form, the emergence of unconventional FFLO phases under the influence of a magnetic field contributes to a similar effect [4,5].

When a spin-singlet type-II superconductor is exposed to a magnetic field, there are two distinct mechanisms which suppress superconductivity and restore the normal metallic state at the upper critical field $H_{c2}$. In most superconductors this is driven by the orbital effect, where the superconducting screening currents reach a pair breaking value, inducing a continuous second-order transition to the normal state [6]. An alternative mechanism is based on the Zeeman effect: when the Zeeman splitting energy between the two electrons of opposite spins, which form the Cooper pair, attains a value that abruptly breaks up all pairs. This occurs at the Pauli limiting field, which is theoretically the maximum magnetic field threshold for superconductivity. This Pauli limit results in a sudden, discontinuous first-order phase transition, restoring the normal state. However, this limit is seldom observed because most spin-singlet superconductors never attain this Pauli limit. In fact, $H_{c2}$ for the majority of type-II superconductors typically occurs at much lower fields and is driven by the orbital effect. In the case of strongly anisotropic layered superconductors, an exception can occur when the magnetic field is applied strictly parallel to the layers [4,7-21]. When the interlayer coupling is weak, the orbital effect is suppressed, allowing the orbital limit to surpass the Pauli limit [22,23]. Typically, the $H_{c2}$ transition line in the magnetic field versus temperature phase diagram then exhibits a steep initial slope, but it plateaus at the Pauli limiting field due to the abrupt suppression of superconductivity by the Zeeman effect.

The Fulde-Ferrell-Larkin-Ovchinnikov (FFLO) state [24,25], an unusual superconducting state, can manifest under specific conditions in type-II spin-singlet superconductors when they exceed the Pauli limit. The formation of the FFLO state provides a solution for the superconductor to preserve its superconducting state even above the Pauli limit. This is achieved by forming Cooper pairs with finite center-of-mass momentum. In this state, the amplitude of the superconducting order parameter undergoes spatial modulation, enabling the superconductor to exist in fields beyond the Pauli limit. To date, it has been observed in very few superconductors, which include organic superconductors [10-20], and the heavy-fermion superconductor $CeCoIn_5$, where the FFLO state coexists with a spin density wave [7-9]. Other instances of this phenomenon have been reported in the iron-based superconductors FeSe [26] and $KFe_2As_2$ [21].

Some TMD materials are intrinsic superconductors, including 2H-$NbSe_2$ and 2H-$NbS_2$ with critical temperatures of 7.2 K and 5.5 K in their bulk form, respectively. TMD superconductors are in principle ideal materials for searching for FFLO states. Their layered structure causes high upper critical fields when the field is applied parallel to the layers, which can exceed (2H-$NbS_2$ [4]) or come close to the Pauli limit (2H-$NbSe_2$ [27]). The FFLO state can be realized if the superconductor is in the clean limit with a long electron mean-free path surpassing the coherence length [28], a condition which is met in TMD materials [29]. Recent studies conducted by some of us, have reported thermodynamic evidence for the FFLO state in 2H-NbS$_2$ [4]. The $H_{c2}$ transition line of 2H-NbS$_2$ was found to surpass the Pauli limit at 10 T and exceeded 15 T. Furthermore, the characteristic upturn of the upper critical field line was observed above the Pauli limit, along with an additional field-induced phase transition line near the Pauli limit. These are indicative of the formation of an FFLO state, with the additional phase transition line representing the transition from the ordinary low-field superconducting phase to the high field FFLO phase. Notably, the additional phase transition line distinguishes the nature of the high field phase from Ising superconductivity, which has been found in TMD superconductors in the 2D limit to provide an alternative route to superconductivity beyond the Pauli limit due to a locking of the electron spins to the direction perpendicular to the basal plane [1-3].

The sister TMD compound, 2H-NbSe$_2$, contrasts in that it approaches, but does exceed the Pauli limit due to its significantly weaker anisotropy [27]. This results in a relatively strong orbital effect that prevents it from surpassing the Pauli limit, thereby excluding the formation of the regular FFLO state. However, a new type of so-called orbital FFLO state [30] has recently been reported for less than approximately 40 nm thick 2D 2H-NbSe$_2$ samples [5]. The interlayer orbital effect, caused by the external magnetic field, can facilitate FFLO formation in an Ising superconductor, where the Zeeman effect is essentially inhibited. This mechanism can establish the FFLO state at field strengths significantly below the Pauli limit. Transport measurements have indicated broken translational and rotational symmetries in the orbital FFLO state, which were interpreted as characteristic signs of finite-momentum Cooper pairings.

In this article, we provide thermodynamic evidence for the existence of an orbital FFLO state in a micro-meter thick 3D bulk single crystalline sample of 2H-NbSe$_2$. We conducted high-resolution DC magnetization and magnetic torque experiments with a magnetic field orientation strictly parallel to the basal plane formed by the NbSe$_2$ layers. See Ref. 4 and 27 for a detailed description of the alignment procedure. As previously reported, the $H_{c2}$ extrapolates to a magnetic field of 11.4 T at zero temperature when the field is aligned within 0.1 degrees with respect to the parallel orientation. This is below the theoretical weak coupling Pauli limit, which can be estimated using the BCS approach as $H_P = 1.85T_c = 13.3$ T. However, when we apply magnetic fields of 3 T and above, the superconducting transition already exhibits weakly first-order behavior at the upper critical field line, which is absent for misalignments on the order of 1 degree. This phenomenon typically indicates that Pauli paramagnetic effects are beginning to exert a pairbreaking influence. In addition, we report here a sharp step-like anomaly appearing in the reversible component of the magnetic torque within the superconducting state. This forms a phase transition line with a slight downward trend upon increasing temperature, clearly indicating the thermodynamic signature of a phase transition separating the ordinary low-field superconducting phase from an unusual high magnetic field state. Furthermore, our precise field-angle resolved electrical transport experiments revealed that the critical field line develops a six-fold symmetry in the field range where this high-field phase occurs. These observations suggest the potential existence of an orbital FFLO state in bulk 2H-NbSe$_2$ samples, making this novel pair density wave state highly accessible in very clean bulk samples and within a temperature and field range that can be accessed in a standard low-temperature laboratory.

# Results

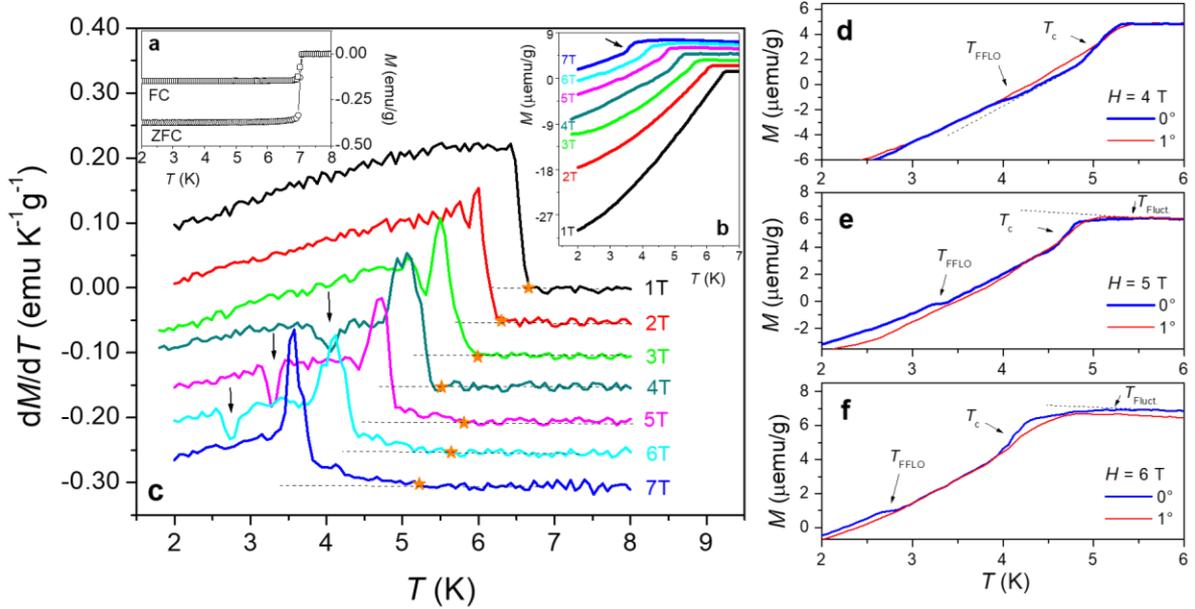

**Fig. 1. DC magnetization data of 2H-NbSe$_2$ in magnetic fields applied parallel to the basal plane.**
a) Low field data measured in a 5 Oe applied magnetic field under zero-field-cooled (ZFC) and field-cooled (FC) conditions. It shows the superconducting transition at $T_c$ = 7.2 K. b) High-field data from 1 – 7 T. The ZFC and FC data was almost identical and only ZFC data is shown. Offsets of 0.025 emu/g have been added for clarity, except for the 1T-data. Main panel (c): temperature derivative of the high field data d$M(T)$/d$T$, showing the development of a first-order transition peak-like anomaly at the superconducting transition in fields above 3 T and small dips at the FFLO transition in 3, 4 & 5 T as marked by the arrows. Stars mark the onset of superconducting fluctuations above $T_c$. Negative offsets of 0.05 emu/K/g have been added for clarity, except for the 1T-data. d – f) Details of the original magnetization data in 4, 5 and 6 T in fields applied strictly parallel to the basal plane (0°) and for a small misalignment (~1°). Step like anomalies appear at the FFLO transition and at $T_c$ only for the 0° data, while for a small misalignment the transition at $T_c$ adopts a more continuous kink-like nature without any FFLO transition.

In Fig. 1 we present DC magnetization data obtained from a 2H-NbSe$_2$ single. Data taken under zero-field-cooled (ZFC) and field-cooled (FC) conditions in a weak 5 Oe applied magnetic field show the zero-field superconducting transition to occur at 7.2 K (Fig. 1a). Measurements in higher magnetic fields up to 7 T were conducted as a function of temperature with a constant field applied parallel to the layered structure (Fig. 1b). At low magnetic fields up to 2 T, we identify a typical kink-like second-order phase transition occurring at the critical temperature. In higher fields (3 T – 7 T) a small step-like feature emerges below the critical temperature (see arrow for the 7T data), indicating a first-order phase transition. The temperature derivative of the magnetization d$M(T)$/d$T$ (main panel, Fig. 1c) reveals more about the phase transition behavior. In the range up to 2 T, the superconducting transition exhibits a step-like behavior, which is the expected behaviour for a thermodynamic quantity being the second temperature derivative of the free energy (like the well-known superconducting specific heat transition at $T_c$). At a magnetic field strength of 3 T, a sharp peak emerges on top of the step-like transition anomaly. As the magnetic field increases further, this peak becomes dominant, which is a characteristic behavior of a first-order transition. The observed field-induced change in the order of the superconducting transition is usually only known to occur in Type-II superconductors due to the Pauli paramagnetic effect, and may indicate that this effect is present here at surprisingly low magnetic fields, given that the theoretical Pauli limit for 2H-NbSe$_2$ with a 7.2K-$T_c$ is roughly estimated to occur near 13 T. A further peculiarity are dip-like transition anomalies occurring in the field range from 4 T to 6 T in d$M(T)$/d$T$, which are also visible in form of an additional tiny step-like transition in the magnetization $M(T)$ as shown in the enlarged data in panels (d) – (f). These steps disappear completely when introducing a small field/basal plane misalignment on the order of 1 degree. We will later identify them as a transition into an orbital FFLO state occurring at characteristic temperatures $T_{FFLO}$. Further details and analysis will be provided later in our discussion. In addition, a small fluctuation tail develops at temperatures above the main superconducting transition in fields above 4 T, as marked by the stars in Fig. 1(c), and by the arrows labelled as $T_{Fluct.}$ in panel (e) and (f).

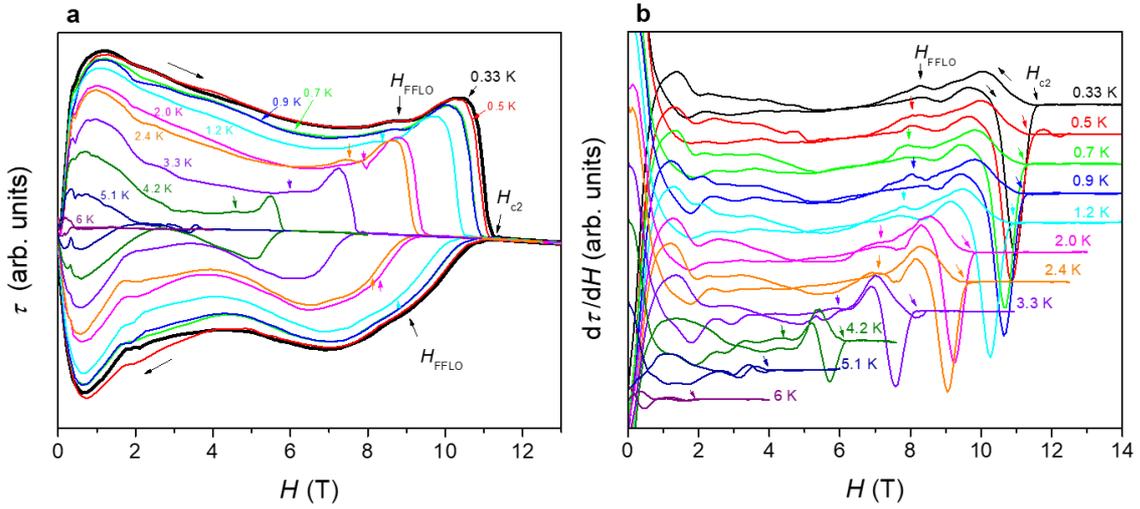

**Fig. 2. Magnetic torque data of 2H-NbSe$_2$ in magnetic fields applied parallel to the basal plane. a** Magnetic torque magnitude $\tau(H)$ measured at fixed temperatures as a function of increasing and decreasing magnetic field (see arrows) applied strictly parallel to the basal plane. $H_{c2}$ is marked by an arrow for the 0.33 K data. Additional arrows indicate the positions of additional small phase transition anomalies attributed to the transition into the orbital FFLO state ($H_{FFLO}$). **b** Magnetic field derivative of the magnetic torque magnitude d$\tau(H)$/d$H$ showing the transition anomalies at $H_{FFLO}$ in both the ascending and descending branches clearer. Arrows mark the transitions at $H_{c2}$ and $H_{FFLO}$. Offsets have been added for clarity.

Fig. 2a shows the magnetic torque for the exactly parallel field orientation at different fixed temperatures. Coming from zero field, the torque signal increases rapidly at small fields towards positive values (see arrow for field sweep direction), reaches a maximum, and then initially decreases. At 5 K the torque continuously disappears at $H_{c2}$ near 4 T, except for a tiny anomaly indicating enhanced flux pinning generally known as the peak effect [32]. At lower temperatures, the torque begins to rise again at higher fields, forming a pronounced peak before abruptly dropping to a small background value in the normal state at $H_{c2}$. Such a sharp decay of the screening currents is usually observed in Pauli-limited superconductors and associated with a first-order nature of the $H_{c2}$ transition triggered by the Pauli paramagnetic effect [4,12,21], thus confirming the findings from the DC magnetization data. Upon decreasing the field, the torque changes sign and the $H_{c2}$ transition appears more continuous, indicating the gradual building up of screening currents in the opposite direction when the field is lowered below $H_{c2}$. In the upward sweep data, small jump-like anomalies appear below the onset of the peak-like anomalies for all temperatures up to 2.4 K (see arrow with label $H_{FFLO}$ marked for the example of the 0.33 K data). We will attribute these in the following to the transition between the ordinary low field superconducting state and the orbital FFLO state in higher fields. We also mark a small change of slope in the 0.33 K data in the downward sweep near 9 T, which we will demonstrate in the following is due to the same phase transition at $H_{FFLO}$.

Fig. 2b shows the first field derivative of the magnetic torque, calculated from the smoothed data in Fig. 2a, which shows the anomalies at $H_{FFLO}$ as small bumps as marked by the vertical arrows. They can be traced up to 2.4 K. Note that the anomaly in the original torque data for decreasing fields is partially hidden by the slope of the broad upper critical field transition, but it is evident here in the field derivative, in which the background slope is transformed into an almost constant contribution.

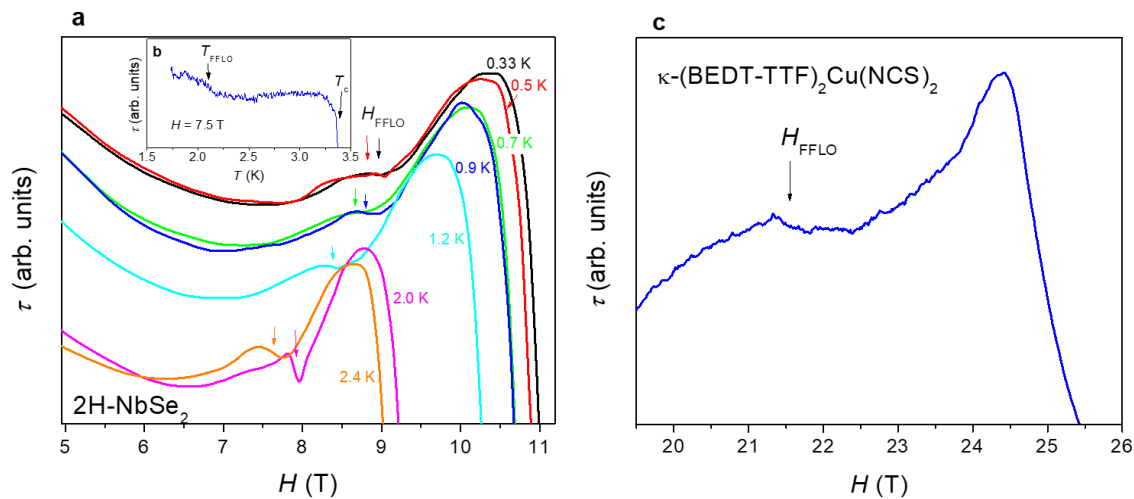

Fig. 3. **Details of magnetic torque data of 2H-NbSe$_2$ in comparison to torque data of the organic FFLO superconductor κ-(BEDT-TTF)$_2$Cu(NCS)$_2$ near the FFLO transition at $H_{FFLO}$. a** Magnetic torque magnitude τ (H) of 2H-NbSe$_2$ measured at fixed temperatures as a function of increasing parallel magnetic field offering an enlarged view on the small step-like anomalies at $H_{FFLO}$. **b** Temperature dependent measurement of the magnetic torque magnitude τ (T) in a parallel 7.5 T magnetic field, which crosses the FFLO transition line at $T_{FFLO}$. **c** Similar enlarged view on the magnetic torque magnitude τ (H) of the layered organic FFLO superconductor κ-(BEDT-TTF)$_2$Cu(NCS)$_2$ as in **(a)**, measured at a fixed temperature of T = 1.7 K as a function of increasing magnetic field applied parallel to the layered structure showing a very similar small step-like anomalies at $H_{FFLO}$ [12].

Fig. 3 shows an enlargement onto the original torque data (**a**) near the anomalies at $H_{FFLO}$ for increasing field, in comparison to torque data on the layered organic FFLO superconductor κ-(BEDT-TTF)$_2$Cu(NCS)$_2$ (**c**) [12]. Both superconductors feature small step-like anomalies at $H_{FFLO}$ which look very similar, suggesting that they may have the same origin. The step occurs in both compounds slightly below a pronounced peak in the torque just before the rapid drop of the signal occurs when approaching the upper critical field. Numerous reports have confirmed the presence of an FFLO state in parallel fields above the Pauli limit at 21 T [11,12,15-18], which occurs in that compound near the Pauli limit of 21 T as shown here by the small step-like anomaly. It is evident that the transition between the ordinary superconducting state at low-fields and the FFLO state at high fields above ~21 T presents a very similar response of the magnetic torque in the form of a small, slightly broadened downward step, as marked by the arrows in Fig. 3a. For 2H-NbSe$_2$, the Pauli paramagnetic effect may already be strong enough to trigger an FFLO state and thus produce such a reversible anomaly in the torque.

The inset Fig. 3b shows temperature dependent magnetic torque data in a 7.5 T magnetic field, which also shows a broad small step-like anomaly, suggesting that this measurement crosses the FFLO transition line as a function of temperature.

In Fig. 4, we plot the extracted $H_{c2}(T)$, $H_{FFLO}(T)$, $T_c(H)$ and $T_{FFLO}(H)$ values of 2H-NbSe$_2$ for the parallel field direction with respect to the basal plane of the layered structure in a magnetic phase diagram. $H_{c2}$ was taken as the upper limit at which the torque deviates from the small normal state background. Although $H_{c2}$ in 2H-NbSe$_2$ does not reach the theoretical Pauli limit, all torque data measured up to 4.2 K show the small reversible torque anomaly, indicating a phase transition within the superconducting state, which is also visible in the temperature dependent measurement of the DC magnetization. While the strong orbital effects naively suggest that no FFLO state should form, even though the paramagnetic effect could have an impact as early as 3 T (suggested by peak-like anomalies in the temperature derivative of the DC magnetisation in Fig. 1 and the rapid decay of screening currents in the torque below $H_{c2}$), it is interesting that this line appears in the phase diagram of 2H-NbSe$_2$. Indeed, such a transition line would be expected during the formation of the FFLO state [4,8,12,21]. We will later discuss that this agrees indeed with the theoretical expectation for the new type of orbital FFLO state in the bulk limit of 2H-NbSe$_2$.

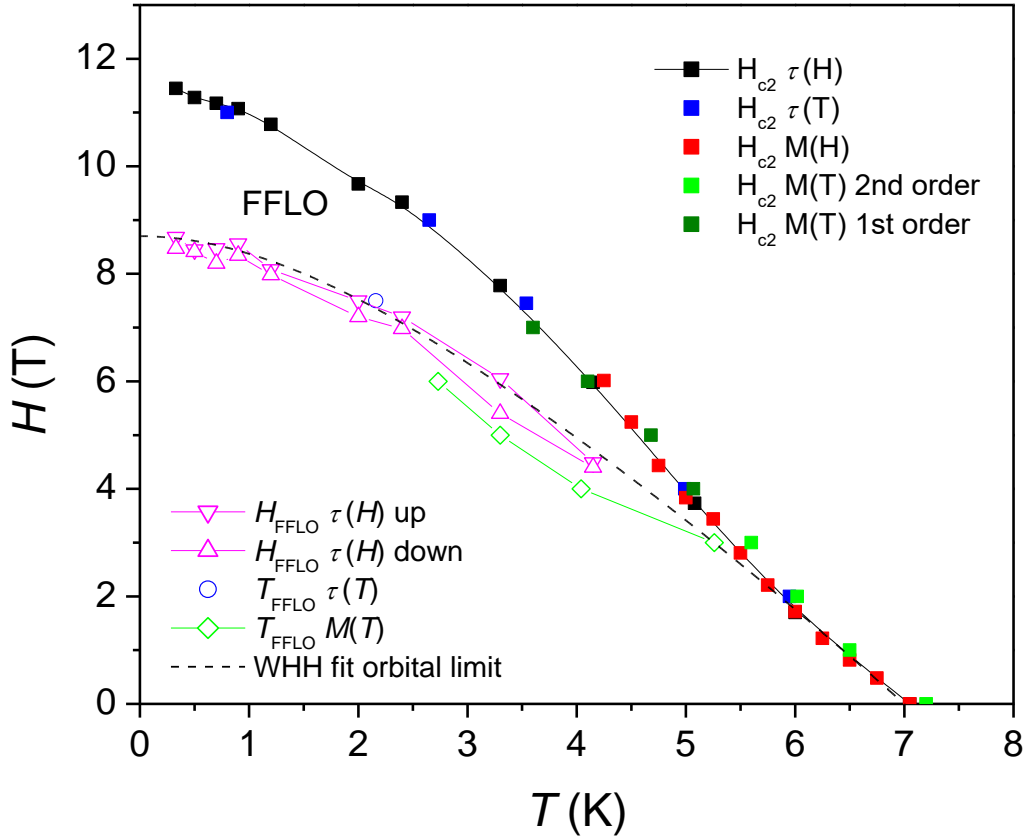

**Fig. 4. Magnetic phase diagram of 2H-NbSe$_2$ in magnetic fields applied strictly parallel and perpendicular to the layer structure**. The data were obtained from magnetic torque measurements and below 7 T by DC magnetization measurements conducted as a function of field ($M(H)$) or temperature ($M(T)$) [27]. The continuous black line marks the $H_{c2}$ line as obtained from the various datasets. Black squares represent $H_{c2}$ data as obtained from the data shown in Fig. 2 ($\tau(H)$). Blue squares have been obtained from temperature dependent magnetic torque measurements $\tau(T)$ [27]. The magenta-coloured triangles mark the small additional reversible anomalies in $\tau(H)$ attributed to a phase transition between the ordinary superconducting state at low fields and the high-field orbital FFLO phase ($H_{FFLO}$), where upright, $\tau(H)$ up, (downright, $\tau(H)$ down) triangles have been measured upon increasing (decreasing) field. The blue circle marks a small downward step-like anomaly in a temperature dependent torque measurement which crosses the FFLO transition line at $T_{FFLO}$ [27]. Green data is obtained from DC magnetization (light green squares: 2$^{nd}$ order transition at $T_c$, dark green 1$^{st}$ order transition). The open diamonds mark the FFLO transition from the $M(T)$ data. We also used a fit of the standard Werthamer–Helfand–Hohenberg (WHH) model (dashed line) for the orbital limit of superconductivity fitted to the data below 2T, which surprisingly stays below the $H_{c2}$ line in higher fields, but perfectly follows the line of the FFLO transition.

Another peculiarity is that the upper critical field line deviates near 3 T from the WHH model which represents the orbital limit, which we fit to the data in the low field range below 2 T in form of an upturn at around 3 T, which is also the field where the FFLO transition line appears to meet the $H_{c2}$ line in form of a tri-critical point. This could indeed represent the characteristic upturn of the $H_{c2}$ line when the orbital FFLO state forms, which stabilizes superconductivity above the orbital limit. In addition, the range of fluctuations gets strongly enhanced above 4 T which appears to be linked to this tri-critical point.

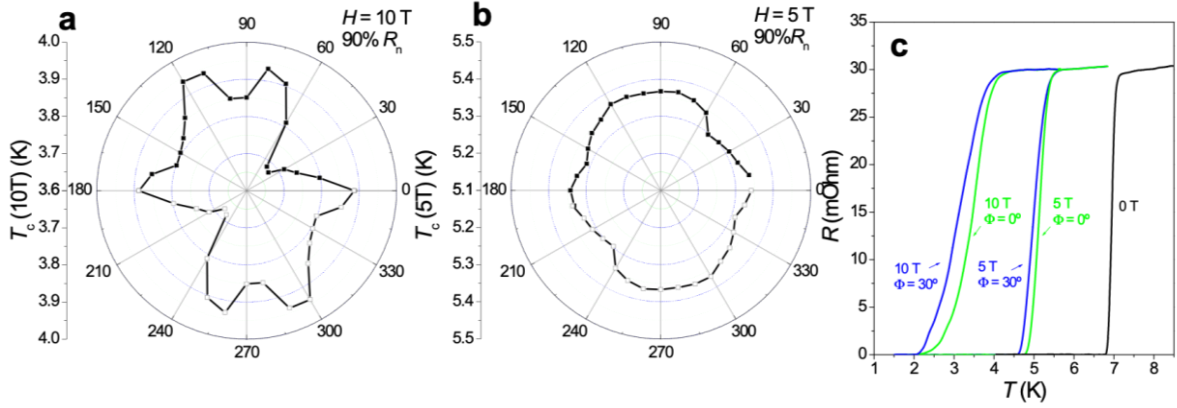

**Fig. 5.** Dependence of the critical temperature on the in-plane orientation of a magnetic field of 10 T (a) and 5 T (b) applied strictly parallel to the basal plane of 2H-NbSe$_2$. A slightly two-fold distorted 6-fold symmetry is seen at 10 T. At 5 T the data is more isotropic. The critical temperature was defined when the resistance reaches 90% of the normal state resistance and the 6-fold symmetry is much less evident. (c) Selected data of the electrical resistance in 0 T and for two different (5 T and 10 T) in-plane magnetic field orientations where a maximum ($\Phi = 0°$) and minimum ($\Phi = 30°$) critical temperature occur.

In Fig. 5 we present polar plots of the critical temperature $T_c$ measured in a 10 T (a) and 5 T (b) magnetic field as a function of the angular direction of the magnetic field within the basal plane of 2H-NbSe$_2$. Note that $T_c$ measured in an applied field is directly related to $H_{c2}$ and the data thus represents the in-plane anisotropy of the upper critical field transition line, which has been demonstrated to reflect the symmetry of the superconducting order parameter [33,34]. The data has been obtained from electrical resistance measurements for many different orientations of the applied magnetic field, but all parallel to the basal plane. Fig. 5c shows resistance data in zero field and at two selected angles of 0° and 30° in 5 T and 10T magnetic field. We used the temperature at which the resistance reaches 90% of the normal state resistance as criterion to determine $T_c(H)$, but very similar results are found for other criteria. Fig. 5a displays a pronounced six-fold symmetry, with maxima approximately each 60 degrees, but with a slight two-fold distortion along the 120 degrees direction. In 5 T the six-fold symmetry is much less evident and almost vanishes. Note that both polar plots are plotted over the same temperature range of 0.4 K, which allows fair comparison of both datasets. The appearance of such a six-fold symmetry has been linked to the orbital FFLO state in 2D 2H-NbSe$_2$ [5], and our data suggests that it is caused by the unusual superconducting phase above $H_{FFLO}$.

**Discussion**

Our data reveals a phase transition line separating the superconducting phase in the magnetic field vs. temperature phase diagram of 2H-NbSe$_2$ into a low field region and a high field region which occurs at low temperatures near 8 T and descends towards higher temperatures down to at least 4 T. The observed field range aligns with the magnetic field interval where first-order transition anomalies manifest during the superconducting (upper critical field) transition in both DC magnetization and magnetic torque. This correspondence suggests the influence of

Pauli paramagnetic pairbreaking. Furthermore, a development of a two-fold distorted 6-fold in-plane symmetry of the upper critical field is observed, which has been reported to be a hallmark of the orbital FFLO phase in very thin flakes of NbSe$_2$ in the 2D limit [5].

The phase transition anomaly occurs in the reversible component of the magnetic torque and is visible both in increasing and decreasing fields with only a marginal hysteresis, as well as in the temperature dependent DC magnetization measurement. Torque is directly related to the anisotropic component of the magnetization, therefore the reversibility of this anomaly hints at a true thermodynamic phase transition as confirmed by its observation in the DC magnetization. This small phase transition at $H_{FFLO}$ as well as the broad peak-like anomaly in the torque between $H_{FFLO}$ and $H_{c2}$ resemble clearly torque data of the layered organic superconductor κ-(BEDT-TTF)$_2$Cu(NCS)$_2$ [12], where the small step-like anomaly and the peak have been shown to originate from the formation of an FFLO state above the Pauli limit. $H_{c2}$ of 2H-NbSe$_2$ remains below the expected Pauli limit, even though it comes near to it, where Pauli paramagnetic effects may already play a role, as evidenced by the magnetization data. The broad pronounced peak effect below $H_{c2}$ further indicates a change in the flux pinning properties between $H_{FFLO}$ and $H_{c2}$, which can be plausibly explained by the formation of a spatially-modulation of the superconducting order parameter. Most importantly, the slope of the upper critical field line begins to deviate from the WHH model of the orbital limit of superconductivity, with an upward trend, in the field region where the FLLO transition line appears to meet the upper critical field line, likely representing the characteristic upturn of the upper critical field line that is considered a hallmark of the FFLO state.

Following the observation of a new type of orbital FFLO state [30, 35] in less than 40 nm thin 2H-NbSe$_2$ samples in Ref. 5 a plausible explanation of our data is thus that this orbital FFLO state exists even in bulk samples. The observation of a thermodynamic phase transition line demonstrates the existence of an unusual magnetic field induced superconducting state. The uniqueness of this phase is its traceability to magnetic fields as low as 4 T and temperatures up to 4 K. The phase transition suggests that the tri-critical point where it intersects with the superconducting transition line could be close to 5 K and fields between 3 and 4 T. Here the $H_{c2}$ transition line indeed shows an upturn and deviates from the WHH model as shown in Fig. 4 [5], which in parallel coincides with the sudden increase in phase incoherent fluctuations above the critical field line.

In the bilayer limit, orbital FFLO states can form under sufficiently high in-plane magnetic fields, with Cooper pairs on the two layers creating a Josephson vortex to screen the field [35]. As the sample thickness increases, the number of Josephson vortices along the out-of-plane direction may also increase. When the thickness is smaller than the Josephson vortex size, only a single Josephson vortex will form, as observed in Ref. 5. For larger thicknesses, multiple Josephson vortices can form, with their inter-vortex distance depending on the field strength. In the bulk limit, Josephson vortices can be closely packed along the out-of-plane direction, resulting in orbital FFLO states throughout the sample [5, 35]. At lower temperatures and higher fields, individual monolayers become decoupled because the Cooper pair momenta of adjacent monolayers are opposite. Consequently, the in-plane upper critical fields can be described by the monolayer limit as the temperature approaches zero.

In Fig. 6, we qualitatively illustrate the three different limits affecting the phase transition lines in the magnetic phase diagram. The blue line and the red dashed line represent the pure orbital effect, with finite Cooper pair momentum **q** (blue) or **q** = 0 (red). These limits may apply to high temperatures and low fields, where Josephson couplings between different layers are significant. As the temperature decreases and the field increases, the Zeeman effect becomes prominent. In this scenario, neighboring layers have opposite Cooper pair momentum and weakly couple to each other. The depairing effect of the field is primarily due to the Zeeman effect within individual layers. This corresponds to the black line, which starts from the zero-field critical temperature of a monolayer, a value smaller than that of the bulk. The interplay of these temperature-dependent critical fields, driven by a dimensional crossover, can qualitatively explain the experimentally observed features in the phase diagram, including the formation of an orbital FFLO state.

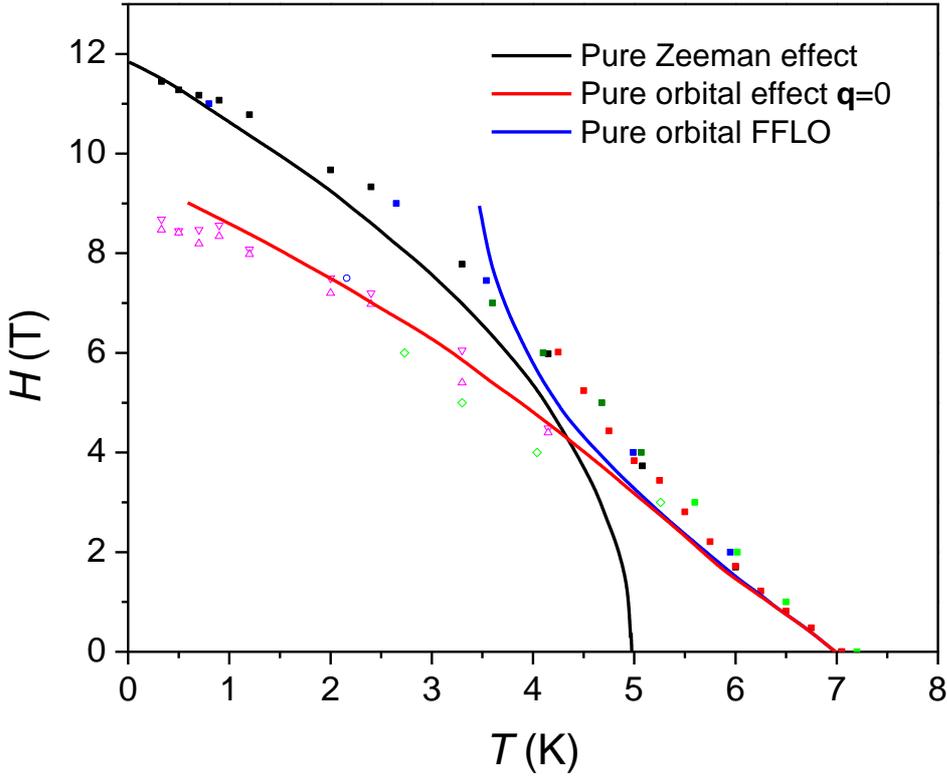

**Fig. 6.** Qualitative illustration of the temperature dependence of critical fields according to theoretical considerations (see Ref. 35 for details). The blue line and the red lines represent the pure orbital effect, with finite Cooper pair momentum **q** (blue) or **q** = 0 (red), expected to be dominant in the high-temperature / low-field limit. The black line represents the low-temperature, high-field limit where the Zeeman effect decouples neighboring layers, making the depairing effect primarily due to the Zeeman effect within individual layers. This line starts from the zero-field critical temperature of a monolayer. The symbols are experimental data identical to those shown in Fig. 4.

In Ref. [27] it was shown that NbSe$_2$ is much less 2D than for example the FFLO superconductor NbS$_2$ [27] with a much weaker anisotropy $\xi_{0\parallel}/\xi_{0\perp}$ = 2.29 in contrast to $\xi_{0\parallel}/\xi_{0\perp}$ = 15 in NbS$_2$. The orbital effect in NbSe$_2$ is thus quite strong, even in parallel magnetic fields, which has been attributed to the ratio of $\xi_{0\perp}$ to the interlayer spacing $d$. The $c$-axis lattice parameter in the out-of-plane direction is about $d$ = 18 Å and the coherence length of NbSe$_2$ $\xi_{0\perp}$ = 31 Å is coupling adjacent layers strongly with strong Meissner currents in the out-of-plane direction, suppressing superconductivity by the orbital effect before the Pauli limit is reached. Nevertheless, it is likely that Ising spin orbit interaction still plays a role in bulk 2H-NbSe$_2$. Since it has been shown that the interlayer orbital effect, caused by the external magnetic field, can facilitate FFLO formation in an Ising superconductor, the FFLO state can indeed be established at field strengths significantly below the Pauli limit [5]. After all, even 30 nm for which the orbital FFLO state has still been observed from upper critical field measurements, is already pretty much in the bulk limit with a large number of layers $N$. Thus, our observed phase transition possibly has the same physical origin as the one described in Ref. 5.

An alternative interpretation could be a vortex melting transition, a phenomenon well-documented in the high-temperature cuprate superconductor YBa$_2$Cu$_3$O$_{7-\delta}$ [36], which has also been considered for 2H-NbSe$_2$ [37-40]. While this explanation is plausible for magnetic fields applied perpendicular to the layers, where an order-disorder transition has been directly observed using scanning tunnelling microscopy [38], it becomes less likely for magnetic field orientations parallel to a strongly layered superconductor. In this case, the vortices exist as Josephson vortices situated between the layers, and the strong pinning effect of the layers would likely inhibit a vortex melting transition or at least push it very close to the upper critical field line. Crucially, our electrical resistance measurements demonstrate that even at magnetic fields as high as 10 T, where only the high-field phase is present, the resistance drops to zero. This unequivocally rules out the possibility of a vortex liquid state, which invariably exhibits a finite electrical resistance. Previous reports of a vortex melting transition for this parallel field orientation [40] could be alternatively explained by the orbital FFLO transition. The formation of an FFLO state, with its spatial modulation of the order parameter, could induce a similar change in the vortex response.

In conclusion, our experimental observations, when coupled with theoretical considerations [35], suggest that the formation of an orbital FFLO state provides a plausible explanation for the observed magnetic field vs. temperature phase diagram of bulk 2H-NbSe$_2$ in strictly parallel magnetic fields. This includes an additional phase transition line within the superconducting phase and a weakly-first-order nature of the upper critical field transition line. This underscores the necessity for a thermodynamic probe to investigate the FFLO transition within the superconducting phase in thicker 2H-NbSe$_2$ samples. This is particularly important as the characteristic upturn of the $H_{c2}$ line appears to become almost hidden when approaching the bulk limit.

## Acknowledgments

We thank U. Lampe and A.T. Lortz for technical assistance. This work was supported by grants from the Research Grants Council of the Hong Kong Special Administrative Region, China (GRF-16303820). M.A.H. expresses gratitude to Alexander Vasiliev for engaging in a fruitful discussion and acknowledges the support received from the VR starting grant 2018-05339. J.P., B.G. and K.P. acknowledge the DGIST institution program (22-BRP-07).

## Methods

For this study, we employed two distinct high-quality 2H-NbSe$_2$ single crystals sourced independently. Remarkably, both crystals exhibited nearly identical superconducting parameters, with a critical temperature of 7.2 K and upper critical field lines in agreement. Sample 1 was used for all magnetic torque measurements, field-dependent DC magnetization measurements and electrical resistivity measurements, while Sample 2 was used for repeated DC magnetization measurements at fixed temperatures. Sample 1 was grown at Uppsala University using an evaporation technique detailed in previous literature [4,41]. The specimen under study was measuring about 1mm x 1mm x 0.1 mm with surfaces that were optically flat. Sample 2 was measuring about 0.4 mm x 0.7 mm x 0.1 mm. It was grown at DGIST using the Se self-flux method. Nb slugs (99.95%, Alfa Aesar, USA) and Se shots (99.999%, Alfa Aesar, USA) were mixed in a molar ratio of 4:96 and loaded into an alumina crucible, which was sealed in a thick quartz ampoule under vacuum. The mixture was heated to 1173 K for 9 hours, maintained at this temperature for 24 hours for complete melting, then cooled to 1123 K for 3 hours, followed by a 10-hour hold. The ampoule was gradually cooled to 1053 K at a rate of 1 K/h and centrifuged, resulting in 2H-NbSe$_2$ single crystals with hexagonal plate morphology.

DC magnetization measurements were conducted at fixed temperatures during field sweeps through the entire hysteresis cycle and under zero-field cooled (ZFC) and field cooled (FC) conditions with a Quantum Desing MPMS 3 VSM SQUID magnetometer in parallel magnetic fields up to 7 T. Since there was no apparent difference in the ZFC and FC data in magnetic field of 1 T or above, only ZFC data is shown for high fields.

The magnetic torque, a vector quantity, is directly associated with the anisotropic component of the DC magnetization **M** and is given by the equation $\boldsymbol{\tau} = \mathbf{M}\times\mathbf{B}$, where **B** represents the applied magnetic field vector. This was measured employing a capacitive cantilever method [4,12,21]. The sensor was installed on a piezo rotator in a 15 T magnet cryostat's $^3$He probe, enabling millidegree precision alignment of the sample's layered structure relative to the field direction, which was achieved by carefully minimizing the torque signal at a fixed field and temperature as described and demonstrated in detail elsewhere [4,12,21,27]. Field or temperature sweeps were conducted at rates of 0.1 - 0.5 T/min and 0.04 K/min, respectively. The 2H-NbSe$_2$ data shown here is identical to data included in Ref. 27, but the publication of Ref. 5 has stimulated us to reinvestigate our data with respect to a possible realization of the orbital FFLO state. The ~100 microns thin sample was attached with highly diluted GE7031 varnish to the flat, polished cantilever plate of a capacitive torque sensor. Capacitance was measured using a General Radio 1615-A capacitance bridge combined with a SR830 digital lock-in amplifier. It should be noted that even at parallel field alignment with respect the basal plane of a superconductor the magnetic torque usually remains finite due to quadrupole magnetic moments [4,12,21]. Data is presented as the magnitude of the torque τ or its field derivative.

Additional field-angle-resolved electrical resistivity measurements were carried out with the sample mounted flat on an Attocube piezo rotator, allowing the applied magnetic field direction to be rotated within the NbSe$_2$ basal plane. Electrical contacts were established using silver-loaded paint. Measurements of the electrical resistance were performed with a Keithley 6221 AC current source in combination with a SR850 digital lock-in amplifier in a constant magnetic field as a function of increasing temperature after field cooling and were repeated for numerous orientations of the in-plane field orientations. The precise parallel sample orientation was ensured with the assistance of an additional goniometer.

# Author Contributions

This work was initiated by R.L., T.T.L and C.S.H. carried out the DC magnetization measurements. C.-w.C. and C.Y. carried out the magnetic torque experiments with help of C.Y.N., K.T.L and T.T.L performed the field angle resolved experiments, A.R.A. and M.A.H. provided the single crystalline sample. The manuscript was prepared by R.L. and all authors were involved in discussions and contributed to the manuscript.